\documentclass[aps,prb,preprint,superscriptaddress,showpacs,floatfix]{revtex4}

\usepackage{graphicx}
\begin{document}

\title{Ferromagnetic Domain Structure of La$_{0.78}$Ca$_{0.22}$MnO$_3$ Single Crystals}

\author{G. Jung}
\affiliation{Department of Physics, Ben Gurion University of the Negev, P.O. BOX 653,
84105 Beer Sheva, Israel}\affiliation{Laboratoire des Solides Irradi\`{e}s, CNRS UMR
7642 \& CEA/DSM/DRECAM, Ecole Polytechnique, 91128 Palaiseau, France}

\author{V. Markovich}
\affiliation{Department of Physics, Ben Gurion University of the Negev, P.O. BOX 653,
84105 Beer Sheva, Israel}

\author{C. J. van der Beek}
\affiliation{Laboratoire des Solides Irradi\`{e}s, CNRS UMR 7642 \& CEA/DSM/DRECAM,
Ecole Polytechnique, 91128 Palaiseau, France}

\author{D. Mogilyansky}
\affiliation{Institute of Applied Research, Ben Gurion University of the Negev,
P.O.Box 653, 84105 Beer Sheva, Israel}

\author{Ya. M. Mukovskii}
\affiliation{Moscow State Steel and Alloys Institute, 119991, Moscow, Russia}

\date{\today}

\begin{abstract}
The magneto-optical technique has been employed to observe spontaneous ferromagnetic
domain structures in La$_{0.78}$Ca$_{0.22}$MnO$_3$ single crystals. The magnetic
domain topology was found to be correlated with the intrinsic twin structure of the
investigated crystals. With decreasing temperature the regular network of
ferromagnetic domains undergoes significant changes resulting in apparent rotation of
the domain walls in the temperature range of 70-150 K. The apparent rotation of the
domain walls can be understood in terms of the Jahn-Teller deformation of the
orthorhombic unit cell, accompanied by additional twinning.
\end{abstract}

\maketitle

\section{Introduction}
The remarkable magnetic and transport properties of the mixed-valence manganese
perovskites of the form $R_{1-x}$A$_x$MnO$_3$, where $R$ is a rare-earth ion and A is
a divalent ion, continuously attract attention of many research groups. The central
feature is a huge decrease in resistivity on application of a magnetic field, referred
to as the colossal magneto-resistance (CMR) effect.\cite{rev,rev1} The properties of
CMR manganites strongly depend on the doping level $x$. In the La$_{1-x}$Ca$_x$MnO$_3$
(LCMO) system the critical doping level $x=x_C=0.225$ separates a nominally
ferromagnetic (FM) insulating and orbitally ordered ground state from a ground state
with a FM metallic character.\cite{rev,rev1,okuda} There is convincing experimental
evidence that the low temperature ground state of the LCMO system within the low
doping range, $0.17 < x < 0.25$, contains distinct FM insulating and FM metallic
phases.\cite{papav,biotteau,PRBmeta} At temperatures below the magnetic ordering
temperature $T_C$, the resistivity of the manganites doped with $x<x_C$ initially
decreases, but with further temperature decrease it increases again.

It is generally agreed that electrical resistance of magnetic domain walls influences
transport properties and magnetoresistance of classical ferromagnets and CMR
manganites.\cite{zeise} Topology of the magnetic domains may therefore play a role in
the temperature dependence of the resistivity even if no consensus has been reached
both on the experimental observability of the domain-wall scattering and on the nature
of the domain wall scattering mechanism.\cite{zeise} The topology of magnetic domains
results from the minimalization of the system free energy and is related to the
magnetic anisotropy of the host material. Surprisingly, still little is known about
the structure and properties of the spontaneous domains in CMR perovskite single
crystals.\cite{domJP}

Magnetic domains in CMR manganites can be efficiently investigated in-situ by means of
the magneto-optical technique (MO). The non-invasive MO allows for monitoring the
evolution of magnetic domains in large areas of the sample as a function of changing
experimental parameters. In our recent report on magneto-optical investigations of
magnetic domains in various LCMO crystals we have shown, among others, that the MO
image of ferromagnetic Weiss domains can be easily confused with the contrast
resulting from twin domains with different magnetic anisotropy.\cite{domJP,domJMMM}

Manganite LCMO crystals are intrinsically twinned in the process of crystallization.
During cool-down they undergo a high temperature phase transition associated with
lowering of the symmetry.\cite{rev,rev1} At room temperature LCMO has an orthorhombic
structure (space group $Pnma$) with lattice parameters $a\approx\sqrt{2}a_p$,
$b\approx 2a_p$, $c\approx\sqrt{2}a_p$, where $a_p$ is the lattice parameter of a
simple cubic perovskite. The long orthorhombic $b$ axis can be directed along any one
of the cubic axis. The $a$ and $c$ orthorhombic axes are perpendicular to $b$ and
rotated $45^{\circ}$ with respect to the cubic axes. Depending on the direction of the
$b$ axis with respect to the cubic cell, three different orientations of the $Pnma$
unit cell are possible. Within each orientation $a$ and $c$ axis can be interchanged,
leading to additional pair of twins. Domains with mutually perpendicular $b$ axes,
refereed to as orthogonal twins, are preferentially separated by \{110\} cubic planes.
Twins with common $b$ axis but with different $a$ and $c$ axis, so-called permutation
twins, are separated by \{100\} cubic planes.\cite{VANAKEN,VANTEN}

In this paper we report on MO investigations of the spontaneous ferromagnetic domain
structure in La$_{0.78}$Ca$_{0.22}$MnO$_3$ single crystals. In particular, we report
on the striking change in the domain topology occurring in the temperature range
70-150 K. The results are interpreted in terms of additional twinning induced by a
possible low temperature Jahn-Teller structural transition.

\section{Experimental and results}
La$_{1-x}$Ca$_{x}$MnO$_3$ crystals were grown by a floating zone method using
radiative heating.\cite{crystal} The X-ray data for the La$_{0.78}$Ca$_{0.22}$MnO$_3$
crystal were compatible with the orthorhombic unit cell $a = 5.4951(5)$ \AA,  $b =
7.7844(6)$ \AA, $c = 5.4947(6)$ \AA, of a perovskite structure. The as-grown crystal
had form of a cylinder, about 4 cm long and 4 mm in diameter. For the MO
investigations we have cut 0.5 mm thick wafers in the direction perpendicular to the
as-grown cylindric crystal axis and polished one of the flat surfaces to optical
quality. Samples for resistive measurements were prepared in the form of small
rectangular bars with vacuum evaporated gold contacts for standard four-point
resistance measurements.

\begin{figure}
\includegraphics*[width=7truecm]{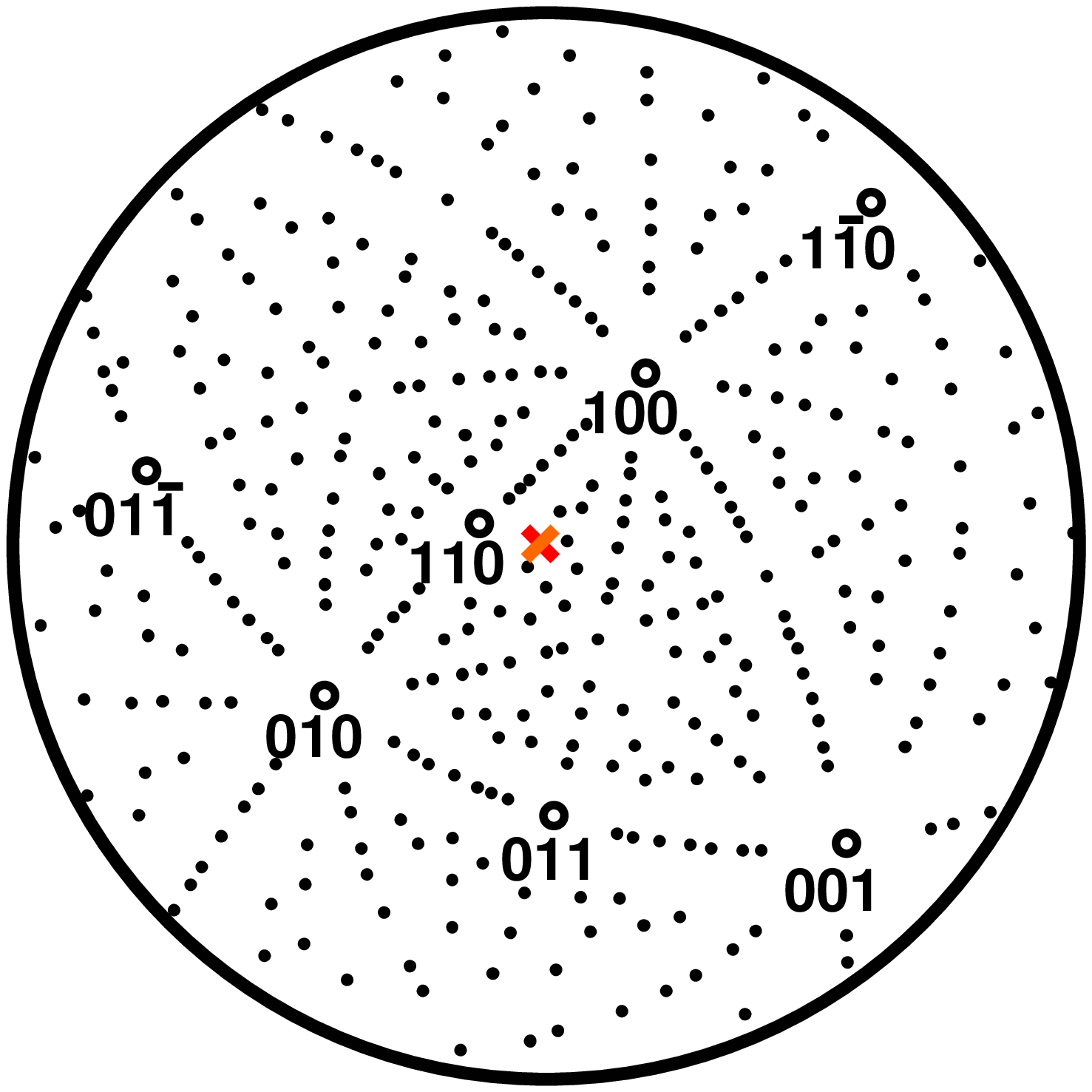}
\caption{The stereographic projection of the wafer plane on which the magnetic domain
structure shown in Fig.~\ref{MO} was observed. The growth direction is indicated with
a cross. The cubic crystallographic planes are marked with circles.} \label{laue}
\end{figure}

The crystallographic orientations of the crystal and wafers were determined by the
Laue method with accuracy of $\pm 2^{\circ}$. The main cylindric axis of the as-grown
crystal was found to deviate 15$^{\circ}$ from the crystallographic [110] cubic
direction. The stereographic projection of the wafer plane is shown in Fig~\ref{laue}.
The growth direction is indicated with a cross and the cubic symmetry crystallographic
planes with circles.

For the magneto-optical imaging a ferrimagnetic garnet indicator film with in-plane
anisotropy was placed directly on the top of the polished surface of the wafer and
observed using linearly polarized light. The reflected light intensity, observed
through an analyzer oriented nearly perpendicularly to the polarization direction of
the incident light, corresponds to the local value of the magnetic induction component
perpendicular to the crystal and to the garnet. To visualize ferromagnetic Weiss
domains the sample was slowly cooled down below $T_C$ in zero applied field. In zero
field conditions the magnetic contrast is not perturbed by magnetic anisotropy
effects, as discussed in details elsewhere.\cite{domJP,domJMMM}

\begin{figure}
\includegraphics*[width=7truecm]{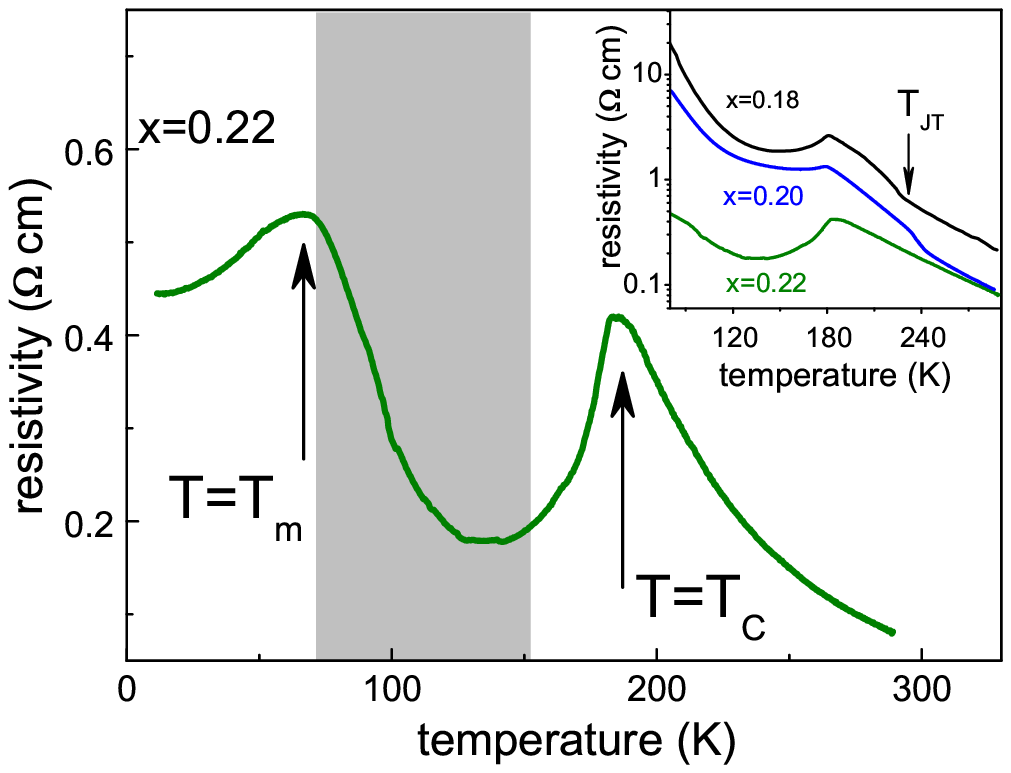}
\caption{Color on-line: Temperature dependence of the resistivity of
La$_{0.78}$Ca$_{0.22}$MnO$_3$ crystal in zero applied magnetic field. The shaded area
indicates the temperature range over which pronounced changes of the magnetic domain
topology is observed. The evolution of the zero field resistivity of the $x=0.22$ LCMO
single crystal is confronted with those doped at $x = $0.18 and 0.20 in the inset.
Note that the step-like change of the resistivity in the paramagnetic phase in
$x=0.18$ and $x=0.20$ crystals cannot be seen in La$_{0.78}$Ca$_{0.22}$MnO$_3$ }
\label{RT}
\end{figure}

Figure \ref{RT} shows the temperature dependence of the zero magnetic field
resistivity of LCMO single crystals with $x$ = 0.18, 0.20, and 0.22. The general
behavior of the resistivity in investigated crystals is the same. With decreasing
temperature the resistivity reaches a pronounced maximum related to the
metal-insulator (M-I) transition at $T=T_{M-I}$. The temperature of the maximum is
very close to Curie temperature determined from independent magnetization
measurements. For La$_{0.78}$Ca$_{0.22}$MnO$_3$ $T_C = 189 \pm 1$ K, while
$T_{M-I}=186$ K.\cite{mark1} At temperatures  below the magnetic ordering temperature
the resistivity decreases in a metallic way with $d\rho/dT> 0$, although the absolute
value of the resistivity is much higher than in common metals. With further
temperature decrease the resistivity exhibits a shallow minimum followed by a strong
low temperature upturn towards a second low-temperature maximum which for $x=0.22$
appears at $T_{m} \approx 68 $ K. The independent magnetization measurements of our
sample did not reveal any peculiarities at temperatures below $T_C$, suggesting that
in the investigated temperature range there is only one magnetic transition at
$T=T_C$.

Double maxima in the $R(T)$ dependence were previously observed in
La$_{2/3}$Ca$_{1/3}$MnO$_{3-d}$ thin films with artificial grain boundaries playing
the role of magnetic tunnel junctions.\cite{gross} The low temperature resistivity
peak in such films was attributed to spin-polarized tunnelling across grain
boundaries. Recently a double peak feature was also observed in optimally doped LCMO
single crystals.\cite{belev} However, in single crystalline samples there are no grain
boundaries and possible barriers for spin-polarized tunnelling may be associated with
domain walls pinned to twin boundaries.\cite{zeise} Twin domain boundaries in CMR
manganites single crystals play a role similar to that of grain boundaries in
polycrystalline samples.\cite{PRBmeta,belev} Due to strong band bending at the
boundary the carrier concentration is depleted and the temperature of the local M-I
transition is depressed.\cite{belev,gross,grossJMMM,Kar} The low temperature peak in
the existing literature is thus interpreted as being due to local M-I transitions
occurring at lower temperature than those in the sample
bulk.\cite{belev,gross,grossJMMM,Kar} Note that the double peak resistivity structure
can be observed only in samples with low specific resistance at low temperatures, such
that relatively low second maximum is not obscured by other resistivity contributions
which strongly increase with decreasing temperature.\cite{okuda}

\begin{figure}
\includegraphics*[width=6truecm]{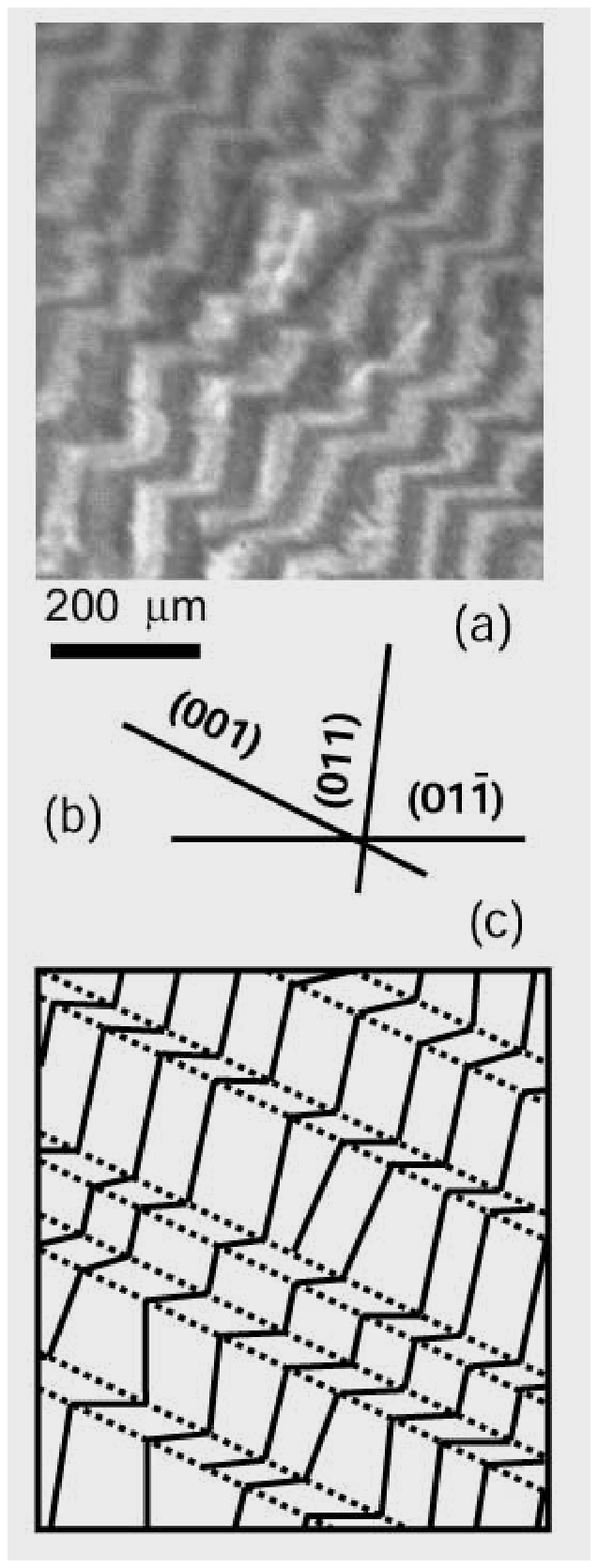} \caption{a) MO image of
the ferromagnetic domains observed in zero field cooled La$_{0.78}$Ca$_{0.22}$MnO$_3$
wafer at 165 K. b) Directions of intersections of the crystallographic planes with the
wafer surface, as determined from the Laue analysis shown in Fig. \ref{laue}. c) The
schematics of the domain structure from the MO shown in a). For clarity of the picture
the lines are drown only along the domains with one direction of the spontaneous
magnetization, which in the MO image appear as bright domains.} \label{schema}
\end{figure}

Figure \ref{schema}a shows a MO micrograph of the polished surface of the $x=22$ LCMO
crystal taken after a slow cooling process down to 160 K in zero applied magnetic
field. As we have pointed out in our earlier paper, these are the conditions under
which unambiguous MO image of the Weiss domains can be obtained.\cite{domJP} The
magnetic contrast associated with ferromagnetic domains appears immediately below
$T_C$. The Weiss domains are  revealed as zigzagging series of dark and bright
stripes. The bright and dark regions in the MO image correspond to areas with opposite
perpendicular component of the stray field, caused by opposite directions of the
spontaneous magnetization in adjacent domains.\cite{domJP} The domains boundaries
follow the intersections of (011) and $(01\bar{1})$ planes with the crystal surface,
as it can be deduced from the schematics shown on Fig.~\ref{schema}c. According to the
crystallographic analysis (011) and $(01\bar{1})$ planes separate structural
orthogonal twin domains in which $b$ axes are mutually perpendicular.\cite{sav} If in
one of the domains in Fig.~\ref{schema} the $b$ axis are directed along the [001]
cubic orientation, then in the neighboring one they will be directed along the [010]
cubic orientation. The applied magnetic field, up  500 Gauss, the maximum field
available in our MO setup, does not change the domains orientation and neither
modifies significantly their width.\cite{domJP} On the basis of the MO investigations
we conclude that magnetic domain walls are pinned to the structural twin domain
boundaries.

\begin{figure}
\includegraphics*[width=\columnwidth]{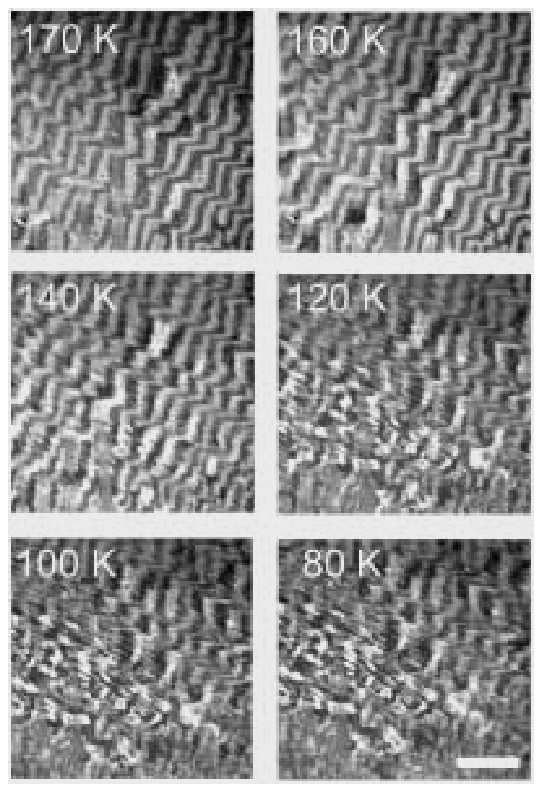}
\caption{Temperature evolution of the ferromagnetic domain structure as revealed by
the MO imaging. The length of the white bar in the bottom right corner is 200 $\mu$m.}
\label{MO}
\end{figure}

Temperature evolution of the shape of ferromagnetic domains in
La$_{0.78}$Ca$_{0.22}$MnO$_3$ crystal is illustrated in Fig.~\ref{MO} showing MO
images of the crystal surface at various temperatures. With temperature decreasing
below $T_C$ the MO contrast increases, and the multi-stripe domain pattern becomes
more and more pronounced. Simultaneously, the sample resistivity decreases due to
increasing strength of the percolation of FM metallic domains, see Fig.~\ref{RT}. The
width of an individual FM domain is in the range of 50 $\mu$m which is about half the
width of the twin domains.\cite{domJP,domJMMM} The normal component of the domain
surface magnetic field is in the range of 40 gauss.\cite{domJP} Such low values of the
domain field, much below the expected saturation magnetization field are, at least
partially, due do the presence of the conical closure domains, clearly seen as
circular spots in the MO micrographs in Fig.~\ref{MO} at temperatures close to $T_C$.

Below 150 K the domain structure starts to undergo significant changes. The smooth
domain walls become corrugated and divide into segments separated by boundaries
parallel to the intersection of the (001) planes with the crystal surface. The new
domains extend in the direction orthogonal to the one dominating at higher
temperatures. The width of the domains extending in the perpendicular direction is
much smaller than the width of the original ones. Eventually, as a result of such
changes, the entire domain structure undergoes apparent rotation and elongates in the
direction perpendicular to the original one. Changes in the domain structure end at
temperatures around 70 K and are reversible. With increasing temperature the dominant
direction of the ferromagnetic domains rotates back to the original one, the
corrugation of the domain walls decreases and eventually disappears. The conical
closure domains reappear in the MO images at temperatures close to $T_C$. However, in
certain restricted areas of the sample, the domains do not fully return to their
pristine state. The shape hysteresis is likely associated with strong pinning of the
modified low temperature domain structure by localized defects. Nevertheless, heating
of the crystal above $T_C$ completely erases the memory of any frozen domain shapes.

\section{Discussion}
Our MO investigations suggest that the magnetic domain walls are pinned to the twin
boundaries. Similar pinning phenomena are common in  ferroelectric and ferromagnetic
systems. Pinning of domain walls by twin boundaries has been observed in ferroelectric
tetragonal perovskites \cite{kalinin} and ferromagnetic alloys with martensitic
structure.\cite{heczko} A surface topography reflecting a twin structure coinciding
with a micromagnetic domain structure was also observed in
La$_{0.83}$Sr$_{0.13}$MnO$_{2.98}$ manganite single crystals.\cite{popov}

Any phase transition associated with lowering of the crystallographic symmetry
generates crystallographic twins in the low-symmetry phase. It is therefore reasonable
to assume that changes in the magnetic domain structure at temperatures below 150 K
are due to additional twinning in the FM phase. The additional twins are directed
perpendicularly to the major twins created at the high temperature cubic to
orthorhombic transition. The appearance of additional perpendicular twin walls in our
sample may be connected with the formation of permutation twins. We suggest that a
possible origin of such excess twinning consists in lowering of the crystal symmetry
by the cooperative Jahn-Teller (JT) transition occurring at low temperatures.

Structural transitions in manganites are invariably accompanied by additional
microstructural strains at surfaces, grain boundaries, or twin planes.\cite{chapman}
The strain fields affect spin orientation and magnetic domains will corrugate and
elongate along the directions of the strain and/or additional twin walls. In
La$_{1-x}$Sr$_x$MnO$_3$ (LSMO) system, which in the low doping regime ($x<x_C$)
behaves very similar to the LCMO, the orthorhombic phase lattice parameters $a$ and
$c$ are almost equal above the temperature of Jahn-Teller coherent distortion
$T_{JT}$. As a result of the JT distortion $a$ and $c$ became significantly different
and the $c/a$ ratio sharply increases.\cite{dabr,geck} Discrepancy between the lattice
parameters leads to significant strain fields which may cause appearance of additional
twins. Orthogonal D2 domains will divide into smaller permutation twins with twin
boundaries directed along the [001] direction. Jahn-Teller transition in low doped
LSMO lowers the crystal symmetry to monoclinic and even triclinic
structure.\cite{geck,cox} A similar transition from the orthorhombic to the monoclinic
phase in Sm$_{0.2}$Ca$_{0.8}$MnO$_3$ crystal was reported to result in the appearance
of additional fine tweed-like domains in the monoclinic phase.\cite{hervieu}

The stoichiometric lanthanum manganite LaMnO$_3$ at room temperature has an
orthorhombic perovskite structure with the $Pnma$ space group symmetry and
antiferrodistorsive orbital ordering (OO) of the Mn-O bond
configuration.\cite{rev,rev1} Alternating long and short Mn-O distances in the
$ac$-plane are the signatures of orbital ordering resulting from the cooperative JT
distortions. Below $T_{JT} \approx 750$ K the JT distortions are static resulting in
O'- orthorhombic axial ratio $b/a < \sqrt{2}$, whereas above $T_{JT}$ they are short
ranged and give pseudocubic ratio $b/a \approx \sqrt{2}$, which is usually designated
as O* - orthorhombic.\cite{rodr} In both LCMO and LSMO systems the temperature of the
O' - O* transition decreases sharply with increasing
$x$.\cite{dabr,geck,klin,asamitsu,liu} The cooperative JT effect is progressively
suppressed, accompanied by developing ferromagnetism.\cite{VANAKEN}  For doping levels
close to $x_C$ the JT transition in both systems appears very close to
$T_C$,\cite{liu} or even below $T_C$.\cite {dabr,geck,klin,asamitsu}

One can see in Fig.~\ref{RT} that the resistivities of $x=0.18$ and $x=0.2$ LCMO
crystals exhibit a step-like feature in the $\rho(T)$ characteristics at
$T_{JT}\approx 230 - 240$ K, commonly regarded as a hallmark of the JT
transition.\cite {liu} Although we do not have a direct proof for the occurrence of
the JT transition in La$_{0.78}$Ca$_{0.22}$MnO$_3$ crystal within the relevant
temperature range, there are several indirect indications confirming such a
hypothesis. Among them, the preliminary extended X-ray absorption fine structure
measurements of our crystal have shown\cite{korea} that Mn-O bond lengths and
Jahn-Teller distortion parameter
$\sigma_{JT}=\sqrt{1/3\sum_i{[(\mbox{Mn-O})_i-<\mbox{Mn-O}>]^2}}$ change with
temperature in a non-monotonic way not only in a close vicinity of $T_C$, as it has
been reported previously for $x=0.25$ LCMO,\cite{radaelli} but also in the temperature
range in which the topology of the magnetic domains changes.

Recent measurements of magnetic, transport, and thermal properties of low doped LCMO
samples have shown that the JT transition in this system likely persists at doping
levels above the percolation threshold, up to $x\approx$ 0.24.\cite{liu} The shift to
higher $x$ in LCMO, with respect to LSMO, is caused by a smaller tolerance factor,
which narrows the $\sigma$-band of electrons. Therefore, one may speculate that JT
transition may also occur in our LCMO sample with $x=0.22$.

The structural JT phase transition in LSMO with $x = 0.14$ occurs in the paramagnetic
phase, $T_{JT}>T_C$, and is of the first order. At slightly higher doping level of $x
= 0.15$, the structural phase transition appears in the ferromagnetic state at
$T_{JT}<T_C$, and becomes of the second order.\cite{klin} A plausible reason for the
absence of the step-like resistivity anomaly in the $\rho(T)$ characteristics of our
sample may be the second order character of the JT transition in $x=0.22$ LCMO.

Monoclinic distortions were found in low-doped LCMO with $x=0.15$ \cite{labanov} and
in the optimally doped, $x\sim 0.3$, epitaxial LCMO films and single
crystals.\cite{belev,lebedev} Recently,  high resolution X-ray diffraction and neutron
powder diffraction structural investigation of low doped LCMO samples have established
that for temperatures below $x$-dependent $T_{JT}$ the symmetry changes from
orthorhombic $Pnma$ to monoclinic $P2_1/c$.\cite{pissas} The transition was directly
revealed through observations of splitting of $(hk0)$ and $(hkl)$ diffraction peaks,
not allowed in the $Pnma$ space group. The JT transition to monoclinic structure was
found to be accompanied by the loss of the mirror plane, signaling a new orbital
ordering with two crystallographically independent Mn sites and a layer-type
arrangement of the different MnO$_6$ octahedra along $b$ axis. For $0.13 < x < 0.175$
the temperature variation of the unit cell volume exhibits a discontinuous change at
$T=T_{JT }$ implying that JT transition is of the first order. However, when $x \geq
0.175$, the magnetic and structural characteristics vary smoothly through
$T_{JT}$.\cite{pissas} This corroborates our assumption that JT structural transition
in $x=0.22$ LCMO is of the second order.

\begin{figure}
\includegraphics*[width=7truecm]{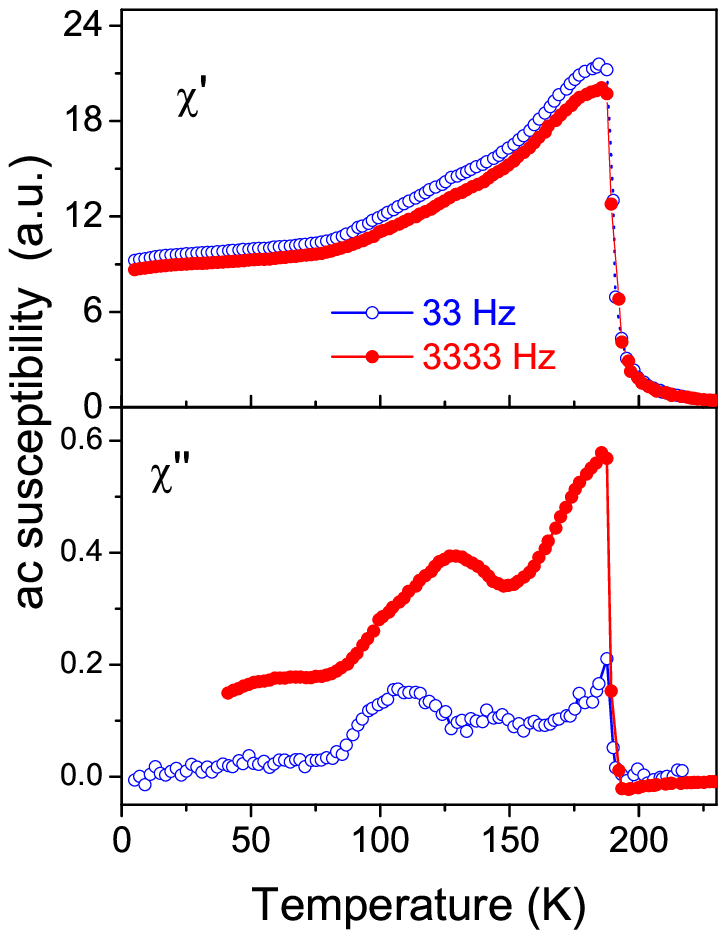}
\caption{Color on-line: The temperature dependence of the ac susceptibility, measured
with 60 mOe ac field, does not exhibit pronounced cluster glass features, as observed
previously in $x=0.18$ and in $x=0.20$ LCMO crystals. The quadrature component of the
ac susceptibility shows a broad smeared maximum around 150 K, which we ascribe to the
motion of additional domain walls excited by the ac field.} \label{chi``}
\end{figure}

The low temperature resistivity upturn in LCMO with $x = 0.18,\; 0.2$ is associated
with strong competition between orbital ordered and orbital disordered ferromagnetic
phases which results in cluster-glass freezing at $T$ $\approx$ 70 K.\cite{rev1,mark1}
The cluster glass-like behavior manifests itself in strong frequency dependence of the
ac susceptibility, remarkable rotation of the easy magnetization axis, and pronounced
difference between zero field cooled and field cooled magnetization curves at the
freezing temperature.\cite{mark1,PRB02}  With an exception of the rotation of an easy
magnetization axis, remarkably less pronounced than in $x = 0.2$ sample, and very
slight frequency dependence of the ac susceptibility, all cluster glass transition
signatures are practically absent in $x$ = 0.22 LCMO crystals.\cite{mark1}
Nevertheless, while the real part of the ac-susceptibility of the
La$_{0.78}$Ca$_{0.22}$MnO$_3$ crystal, see Fig.~\ref{chi``}, does not show any spin
glass peculiarities, the imaginary part $\chi^{''}$ exhibits a broad frequency
dependent maximum in the temperature range 70 -140 K.  The excess temperature
dependent dissipation can be attributed to the motion of magnetic domain walls excited
by the measuring ac magnetic field. Consistently with the scenario proposed in this
paper, additional twinning leads to significant increase in the total length of the
domain walls, what results in a broad $\chi^{''}$ peak.

The experiments show that the temperature range in which the domain topology changes
coincides with the temperature range of the resistivity upturn, see Fig. \ref{RT}. Can
the low temperature resistivity be explained by the observed domain rotation? This is
doubtlessly a very attractive explanation. Nevertheless, one has to remember that a
very similar behavior of the resistivity has been also observed in slightly less doped
LCMO crystals with $x=0.18$ and $x=0.2$,\cite{PRBmeta,PRB02} in which the structural
J-T transition occurs solely in the paramagnetic phase, $T_{JT}>T_C$.\cite{rev1} We
have associated the domain rotation with the structural J-T transition. Consistently,
in these crystals, we have not detected any significant changes in the domain topology
with changing temperature. Therefore, one has to conclude that even if the corrugated
domain walls contribute to the observed resistivity upturn they cannot be accountable
for the entire low temperature behavior of the resistivity which is likely due to
action several interdependent factors.

In summary, transport measurements  and magneto-optical imaging have been employed to
study La$_{0.78}$Ca$_{0.22}$MnO$_3$ single crystal having the Ca-doping level close to
the percolation threshold $x_C$. A prominent spontaneous magnetic domain structure in
the form of zigzagging parallel stripes, with opposite direction of the spontaneous
magnetization in adjacent strips, appears in MO micrographs just below the Curie
temperature $T_C= 189$ K. In the temperature range 70--150 K, corresponding to the low
temperature resistivity upturn, a topological change in the domain structure occurs.
The domain walls corrugate and eventually the entire domain structure rotates
90$^{\circ}$ with respect to the original direction. The evolution of the magnetic
domain structure can be understood in terms of additional perpendicular twinning
associated with a reduction of the crystal symmetry, from orthorhombic to monoclinic,
due to a presumed low temperature cooperative Jahn-Teller transition.

\acknowledgments
This research was supported by the Israeli Science Foundation
administered by the Israel Academy of Sciences and Humanities (grant 209/01) and by
French-Israeli Arc-en-Ciel exchange program. Ya. M. M. acknowledges the support
obtained from ISTC grant 1859.

\end{document}